\documentclass[final]{iau}
\usepackage{graphicx}
\usepackage{natbib}

\newcommand{\apjs}{{ ApJS }}

\newcommand{\mnras}{{ MNRAS }}

\newcommand{\apj}{ApJ}
\newcommand{\aap}{A\&A}
\newcommand{\apjl}{ApJL}

\title[Gravitationally unstable planetary gaps]
      {Gravitational instability of planetary gaps and its effect on orbital
        migration}
      
\author[Min-Kai Lin \& Ryan Cloutier]  
{Min-Kai Lin
 \and Ryan Cloutier}

\affiliation{Canadian Institute for Theoretical Astrophysics, 60
  St. George Street, Toronto, ON, M5S 3H8, Canada \\email: {\tt
    mklin924@cita.utoronto.ca}, {\tt cloutier@cita.utoronto.ca}} 

\pubyear{2013}
\volume{299}  
\jname{Exploring the formation and evolution of planetary systems}
\editors{M. Booth, B. C. Matthews \& J. R. Graham, eds}
\begin{document}

\maketitle

\begin{abstract}
  Gap formation by giant planets in self-gravitating disks may lead to
  a \emph{gravitational edge instability} (GEI). We demonstrate this GEI 
  with global 3D and 2D self-gravitating disk-planet simulations using the 
  ZEUS, PLUTO and FARGO hydrodynamic codes. High resolution 2D 
  simulations show that an unstable outer gap edge can lead to outward
  migration. 
  Our results have important implications for theories of
  giant planet formation in massive disks.  
\end{abstract}

\firstsection
\section{Introduction: the gravitational edge instability}
Models of giant planet formation in self-gravitating
protoplanetary disks may involve gap-opening in massive
disks. However, gaps or edges in self-gravitating disks are
potentially unstable to a gravitational edge instability (GEI)
associated with local potential vorticity maxima
\citep{meschiari08,lin11b}.    

We confirm the GEI for planet-induced gaps with global 
disk-planet simulations. We consider a locally isothermal disk with
constant aspect-ratio $h=0.05$ and Keplerian Toomre parameter
$Q_k=1.5$ at the outer disk boundary $r_\mathrm{out}=2.5r_p$, where
$r_p$ is the fixed orbital radius of a planet with mass
$M_p=10^{-3}M_*$ and $M_*$ is the central star mass. 
We simulated this self-gravitating disk-planet system in 3D using
the ZEUS-MP code \citep{hayes06} and PLUTO code \citep{mignone07}; and
in 2D using the FARGO code \citep{baruteau08}. The 3D simulations are
performed in spherical polar co-ordinates covering two vertical
scale-heights.

\begin{figure}
  \centering
  \includegraphics[scale=0.42,clip=true,trim=0cm 1cm 0cm
    1cm]{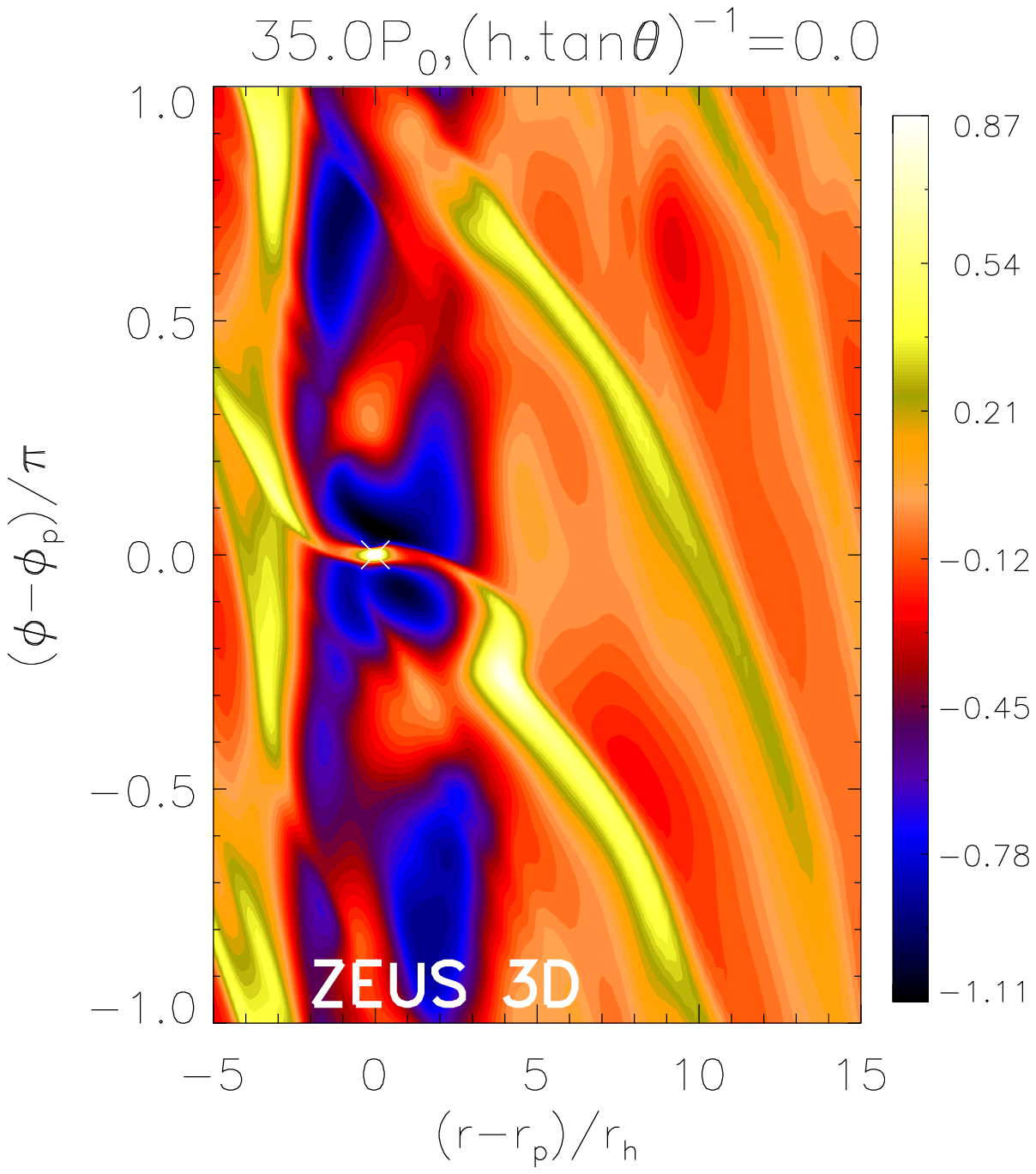}\includegraphics[scale=0.42,clip=true,trim=2.24cm 1cm 0cm
    1cm]{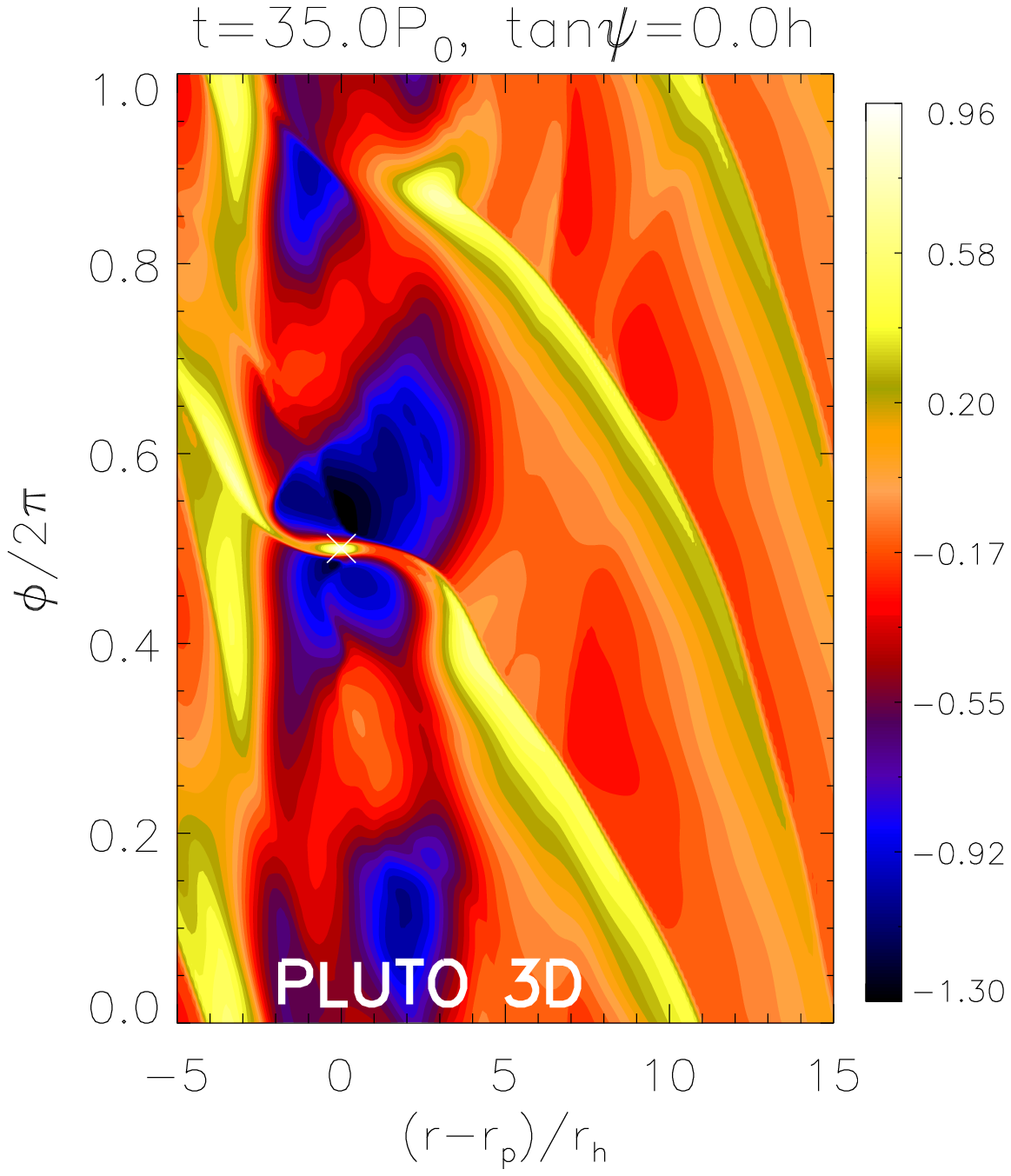}\includegraphics[scale=0.42,clip=true,trim=2.24cm 1cm 0cm
    1cm]{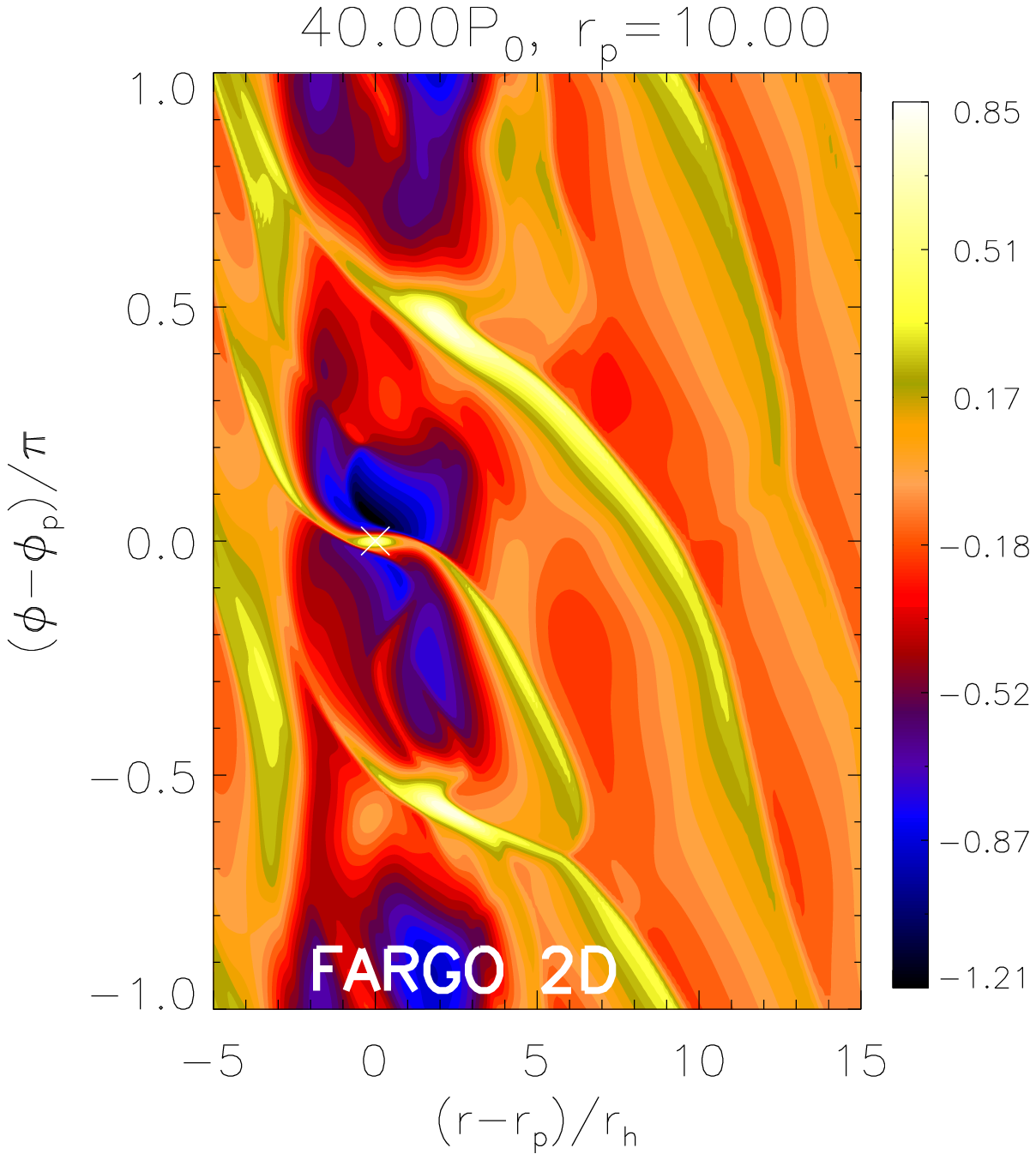}
  \caption{Gravitational edge instability of
    a planetary gap. The logarithmic midplane density perturbation is
    shown for 3D runs (left, middle) and the 
    logarithmic surface density perturbation is shown for the 2D run
    (right). Snapshots are taken at 15 (10) orbits after the planet
    is introduced into the 3D (2D) disk. Horizontal axis are in units
    of Hill radii. The planet is at the origin. The left and right
    simulations are taken from \cite{lin12b} and \cite{cloutier13},
    respectively. \label{gei_demo} 
  }
\end{figure}

Fig. \ref{gei_demo} shows the outer gap
edge develops a spiral instability with azimuthal wavenumber $m=2$
(for 3D runs one of the spiral coincides with the planet wake in the
snapshot). The spirals extend close to $r_p$. In fact, the GEI
co-rotation radius is just inside a giant planet's co-orbital region
\citep{lin11b}. So the GEI supplies fluid onto horseshoe orbits ahead
of the planet, which provides a positive disk-on-planet torque
\citep{lin12}.    

\section{Instability-induced outward migration and implications}
\begin{figure}
  \centering
  \includegraphics[scale=0.4]{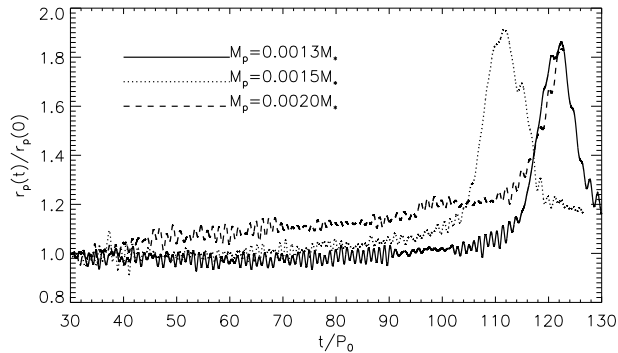}
  \caption{Orbital migration of giant planets when the outer gap edge
    is gravitationally unstable. Here, $P_0$ is the orbital period at the planet's
    initial orbital radius. Adapted from 2D FARGO
    simulations presented in \cite{cloutier13}. \label{migrate}  
  }
\end{figure}

We also performed 2D FARGO simulations with different $M_p$ where the
planet is allowed to migrate, 
so that $r_p=r_p(t)$. Fig. \ref{migrate} shows that for $t\lesssim
100$ orbits, outward migration proceeds at a faster rate with
increasing $M_p$, because the GEI becomes stronger. 
Eventually, a GEI
spiral-planet scattering event triggers rapid outward type III migration.  
In the context of the disk fragmentation model for wide-orbit giant
planet formation \citep{voro13}, this suggests that gap-opening may not be sufficient to
maintain giant planets on a fixed orbital radius beyond a few tens of
orbits.  

\emph{Acknowledgments.}
Several computations were performed on
the GPC supercomputer at the SciNet HPC Consortium. SciNet is funded by: the
Canada Foundation for Innovation under the auspices of Compute Canada;
the Government of Ontario; Ontario Research Fund - Research
Excellence; and the University of Toronto.

\end{document}